\documentclass[sigconf]{acmart}

\usepackage{booktabs}
\usepackage{algorithm}
\usepackage{amsmath}
\usepackage[noend]{algpseudocode}
\usepackage{multirow}
\usepackage{makecell}
\usepackage{scalerel}

\usepackage{tabulary}
\newcolumntype{C}[1]{>{\centering\arraybackslash}p{#1}}
\usepackage{bm}

\usepackage{setspace}

\usepackage{enumitem}
\setlist{nolistsep}

\DeclareMathOperator*{\concat}{\text{\Large $\parallel$}}

\copyrightyear{2021}
\acmYear{2021}
\setcopyright{acmcopyright}\acmConference[KDD '21]{Proceedings of the 27th ACM SIGKDD Conference on Knowledge Discovery and Data Mining}{August 14--18, 2021}{Virtual Event, Singapore}
\acmBooktitle{Proceedings of the 27th ACM SIGKDD Conference on Knowledge Discovery and Data Mining (KDD '21), August 14--18, 2021, Virtual Event, Singapore} \acmPrice{15.00}
\acmDOI{10.1145/3447548.3467220}
\acmISBN{978-1-4503-8332-5/21/08}

\begin{document}
\fancyhead{}

\title{Learning Elastic Embeddings for Customizing\\On-Device Recommenders}
\author{Tong Chen}
\affiliation{%
  \institution{The University of Queensland}
}
\email{tong.chen@uq.edu.au}

\author{Hongzhi Yin}
\authornote{Corresponding author; contributing equally with the first author.}
\affiliation{%
  \institution{The University of Queensland}
}
\email{h.yin1@uq.edu.au}

\author{Yujia Zheng}
\affiliation{%
  \institution{Carnegie Mellon University}
}
\email{yjzheng19@gmail.com}

\author{Zi Huang}
\affiliation{%
  \institution{The University of Queensland}
}
\email{huang@itee.uq.edu.au}

\author{Yang Wang}
\affiliation{
  \city{Hefei University of Technology}
}
\email{yangwang@hfut.edu.cn}

\author{Meng Wang}
\affiliation{
  \city{Hefei University of Technology}
}
\email{eric.mengwang@gmail.com}


\begin{abstract}
In today's context, deploying data-driven services like recommendation on edge devices instead of cloud servers becomes increasingly attractive due to privacy and network latency concerns. A common practice in building compact on-device recommender systems is to compress their embeddings which are normally the cause of excessive parameterization. However, despite the vast variety of devices and their associated memory constraints, existing memory-efficient recommender systems are only specialized for a fixed memory budget in every design and training life cycle, where a new model has to be retrained to obtain the optimal performance while adapting to a smaller/larger memory budget. In this paper, we present a novel lightweight recommendation paradigm that allows a well-trained recommender to be customized for arbitrary device-specific memory constraints without retraining. The core idea is to compose elastic embeddings for each item, where an elastic embedding is the concatenation of a set of embedding blocks that are carefully chosen by an automated search function. Correspondingly, we propose an innovative approach, namely \underline{r}ecommendation with \underline{u}niversally \underline{l}earned \underline{e}lastic embeddings (RULE). To ensure the expressiveness of all candidate embedding blocks, RULE enforces a diversity-driven regularization when learning different embedding blocks. Then, a performance estimator-based evolutionary search function is designed, allowing for efficient specialization of elastic embeddings under any memory constraint for on-device recommendation. Extensive experiments on real-world datasets reveal the superior performance of RULE under tight memory budgets.
\end{abstract}



\maketitle

\vspace{-0.3cm}
\section{Introduction}\label{sec:intro}
Recommender systems are now an indispensable component in online services, providing users with tailored experience based on their preferences. With the fast pace of digitalization and hardware revolution, there has been a recent surge of moving data analytics from cloud servers to edge devices \cite{shi2016edge} to ensure timeliness and privacy. As recommendation services commonly require access to users' personal data, on-device recommendation appears to be an ongoing trend, which facilitates new applications like mobile recommendation \cite{wang2020next} and federated recommendation \cite{muhammad2020fedfast}.

In this regard, the study on memory-efficient recommender systems \cite{lian2020lightrec,wang2020next,joglekar2020neural,shi2020compositional} emerges, where the resulted models are lightly parameterized and can be deployed on resource-constrained edge devices. Notably, the majority of those work is around latent factor models (e.g., matrix factorization), as they exhibit dominant recommendation accuracy and are less dependent on auxiliary side information. In a general sense, latent factor-based recommenders map each user and item into vector representations (i.e., embeddings), then estimate the pairwise user-item affinity via similarity metrics (e.g., vector dot product) or carefully designed deep neural networks \cite{chen2019air,zhang2019inferring,guo2021hierarchical}. Due to the sheer volume of categorical features\footnote{We will mainly refer to users and items in our paper for simplicity.} such as user/item IDs and product types, the embeddings in latent factor models are the main source of memory consumption \cite{wang2020next} rather than other parameters (i.e., weights and biases). Hence, most efforts are made to compress the embeddings \cite{shi2020compositional,lian2020lightrec,liu2020automated} to reduce the size of a recommendation model. 

Early memory-efficient recommenders employ discrete hashing to convert real-valued embeddings into compact binary codes \cite{zhang2016discrete,zhou2012learning}, with which user-item similarities are preserved in the Hamming space. However, a widely acknowledged fact \cite{liu2018discrete,zhang2016discrete} is that, the expressiveness of the resulted binary codes is greatly impaired due to inevitable information loss in this quantization process, leading to constantly inferior recommendation performance even compared with the plain matrix factorization. In light of this, alternatives that generate informative real-valued embeddings with a minimal amount of parameters are recently developed, where two major branches are recommenders based on compositional embeddings \cite{shi2020compositional,lian2020lightrec} and multidimensional embeddings \cite{liu2020automated,zhao2020memory}. Compositional embedding-based recommenders consist of a small number (substantially smaller than the number of all users and items) of meta-embeddings, such that a user/item can be represented as a distinct combination of selected meta-embeddings. Meanwhile, multidimensional embeddings allow for each embedding to have multiple candidate dimensions during training. As each user/item does not necessarily need the largest embedding dimension to encode all its information (e.g., long-tail items with limited predictive signals), the optimal embedding dimension for each user/item can be automatically learned to reduce excessive parameters while retaining the recommendation performance. 

With the increasing diversity of IoT facilities, on-device recommendation is no longer exclusive to cellphones, and rather, becomes feasible on smaller devices such as smart watches, Internet TV boxes, and routers. Heterogeneity exists not only across different device categories, but also across different device ages, e.g., iPhone 11 has doubled the RAM size of iPhone 8 in just two years. Ideally, for the same recommendation task, a model should be customized for each specific device, such that the recommendation performance and utilization of space are both optimized. Unfortunately, in existing paradigms, a memory-efficient recommender specialized for each on-device memory budget has to be designed and trained from scratch, incurring high engineering costs. Though recent advances in automated machine learning (AutoML) can ease the difficulties in the design phase \cite{liu2020automated} with neural architecture search (NAS), retraining is still mandatory every time a new memory constraint needs to be met. The model training and architecture search are normally entangled in NAS-based models, making the learned recommender parameters (i.e., embeddings in our case) and architecture only optimized towards one specific memory budget. As a result, such ``train once, get one'' scheme is too inefficient and unsustainable to keep up with the prosperity of on-device computing, e.g., real-time download requests on a recommender system in app stores from different devices. 

To this end, we suggest a new lightweight, once-for-all recommendation paradigm that is able to automatically adapt to heterogeneous devices with minimal hassles after being fully trained. This is achieved by decoupling the embedding training and search phases via our notion of \textbf{\textit{elastic embeddings}} for on-device recommendation. It is worth noting that in practice, a user's device can only access and store her/his own user embedding, so we will specifically target on compressing all item embeddings in our work. Essentially, instead of learning a single embedding with relatively high dimensionality (e.g., 128-dimensional) for every item, we view a full item embedding as the concatenation of smaller embeddings (e.g., 16 8-dimensional vectors) which we term \textbf{\textit{embedding blocks}}. This is distinct from compositional embeddings~\cite{shi2020compositional} as each item's embedding blocks are exclusive and not shared, ensuring stronger expressiveness. After all embedding blocks are trained, we aim to build an automated search function to identify an appropriate subset of embedding blocks to be concatenated into an elastic embedding for each item. All searched elastic embeddings are then used for on-device recommendation, providing optimal performance under any given memory constraint. 
Elastic embeddings offer abundant flexibility on which and how many embedding blocks can be chosen, controlling the information carried and memory consumed by each item embedding. With an efficient search function, the search cost is negligible compared with retraining a new model, so we can effortlessly configure the most suitable elastic embeddings for heterogeneous devices without retraining. 

However, non-trivial obstacles need to be tackled before we can benefit from this paradigm. Given the large search space spanning across all items' embedding blocks, it is impractical for the search function to enumerate all embedding block combinations and evaluate their performance. Also, as all embedding blocks are trained before on-device deployment, we need to ensure that every embedding block is sufficiently informative to guarantee high-quality recommendations whenever it is selected in the search phase.  
To address those issues, we present a novel solution for \underline{r}ecommendation with \underline{u}niversally \underline{l}earned \underline{e}lastic embeddings, or \textbf{RULE} for short. Specifically, to facilitate efficient yet accurate search for elastic embeddings, we propose to identify suitable embedding blocks at the item group level rather than the individual item level. Intuitively, this allows intra-group items to share similar elastic embedding structures and significantly shrinks the search space. Meanwhile, when training all embedding blocks, we propose a diversity-driven regularizer to encourage variety in the information encoded in different embedding blocks, thus ensuring the expressiveness of arbitrary elastic embeddings. With the trained embedding blocks, we design a performance estimator-based method for evolutionary search \cite{real2019regularized} to bypass the need for evaluating every candidate elastic embedding structure. To further speed up the search process, we introduce a Gaussian prior when determining the number of embedding blocks assigned to each group, so that the search function tend to favor a large/small embedding dimension for only a few really important/unimportant item groups while keeping other groups' embedding dimension balanced. As such, RULE is a low-cost solution to customizing on-device recommenders.

Our contributions in this work are three-fold:
\begin{itemize}
	\item We point out the inflexibility of the current lightweight recommendation paradigm when facing diverse on-device memory budgets, and propose a new, once-for-all recommendation paradigm using elastic embeddings. Our paradigm allows a recommender to be quickly specialized for arbitrary on-device memory constraints without retraining. 
	\item We propose RULE, a novel solution that addresses the key challenges in the execution of this new paradigm. RULE can efficiently search adequate elastic embedding structures to achieve optimal on-device recommendation performance.
	\item We conduct extensive experiments on two real-world datasets under multiple memory constraints. Experimental results show that RULE yields superior recommendation performance compared with state-of-the-art baselines. 
\end{itemize}

\vspace{-0.5cm}
\section{Preliminaries}\label{sec:pre}
We hereby define some key concepts and data structures in our method, namely the \textbf{\textit{embedding block}} and \textbf{\textit{elastic embedding}}.

\noindent\textbf{Definition 1} (Embedding Block): For an arbitrary item $v_j\in \mathcal{V}$, its full embedding $\textbf{v}_j \in \mathbb{R}^{D}$ is the concatenation of $N$ separate $d$-dimensional vectors, i.e., $D=Nd$ and $\textbf{v}_j = [\textbf{e}_{j(1)}, \textbf{e}_{j(2)},...,\textbf{e}_{j(N)}]$ where $\textbf{e}_{j(n)}\!\in\! \mathbb{R}^{d}$ ($1\!\leq n\!\leq N$) is referred to as an \textit{\textbf{embedding block}}. 

\noindent\textbf{Definition 2} (Elastic Embedding): An elastic embedding $\textbf{v}_j^*$ for item $v_j \in \mathcal{V}$ is the concatenation of a non-overlapping subset of its embedding blocks. Mathematically, $\textbf{v}^*_j = \concat_{n\in \mathcal{B}}\textbf{e}_{j(n)} \in \mathbb{R}^{|\mathcal{B}|d}$, where we use $\concat$ to denote the iterative concatenation of vectors, and $\mathcal{B}$ stores the selected block indexes ($|\mathcal{B}|\leq N$). Hence, the dimensionality and expressiveness of $\textbf{v}_j^*$ are both dependant on the composition of $\mathcal{B}$.

\vspace{-0.3cm}
\section{Defining The Search Space of RULE}

Following Section \ref{sec:pre}, at this stage we have $|\mathcal{V}|\times N$ different embedding blocks for all items in total. Assuming each $\textbf{v}^*_j$ receives at least one embedding block, and all embedding blocks in $\textbf{v}^*_j$ are indexed in an ascending order, then the total number of possible combinations is $2^{N}\!-\!1$ for each item, and is $(2^{N}\!-\!1)^{|\mathcal{V}|}$ for all. As the number of items easily reaches millions in modern e-commerce applications, it is impractical to define the search space over $|\mathcal{V}|\times N$ embedding blocks due to the potentially high search cost. Hence, it makes sense for us to narrow down the search space by grouping multiple items together. That is, we segment the item set into $G$ subsets (i.e., groups) via $\mathcal{V}\!=\!\{\mathcal{V}_1, \mathcal{V}_2, ..., \mathcal{V}_G\}$ that are equally sized, i.e., $|\mathcal{V}_g|\!=\!\!\frac{|\!\mathcal{V}\!|}{G}$ for $1\!\!\leq g\!\leq G$. For all items in each group $\mathcal{V}_g$, we concatenate their $n$-th embedding block into a larger, $|\mathcal{V}_g|\times d$-dimensional embedding block\footnote{Directly concatenating $|\mathcal{V}_g|$ column vectors will lead to a $d\times |\mathcal{V}_g|$ matrix but we follow the commonly used look-up table style to keep the notations intuitive.} $\textbf{E}_{g(n)}$. Then the selection of embedding blocks becomes a group-level process, making all items within group $\mathcal{V}_g$ have the same embedding block indexes, denoted by $\mathcal{B}_g$.

\begin{figure}[!t]
\center
\includegraphics[width = 3.3in]{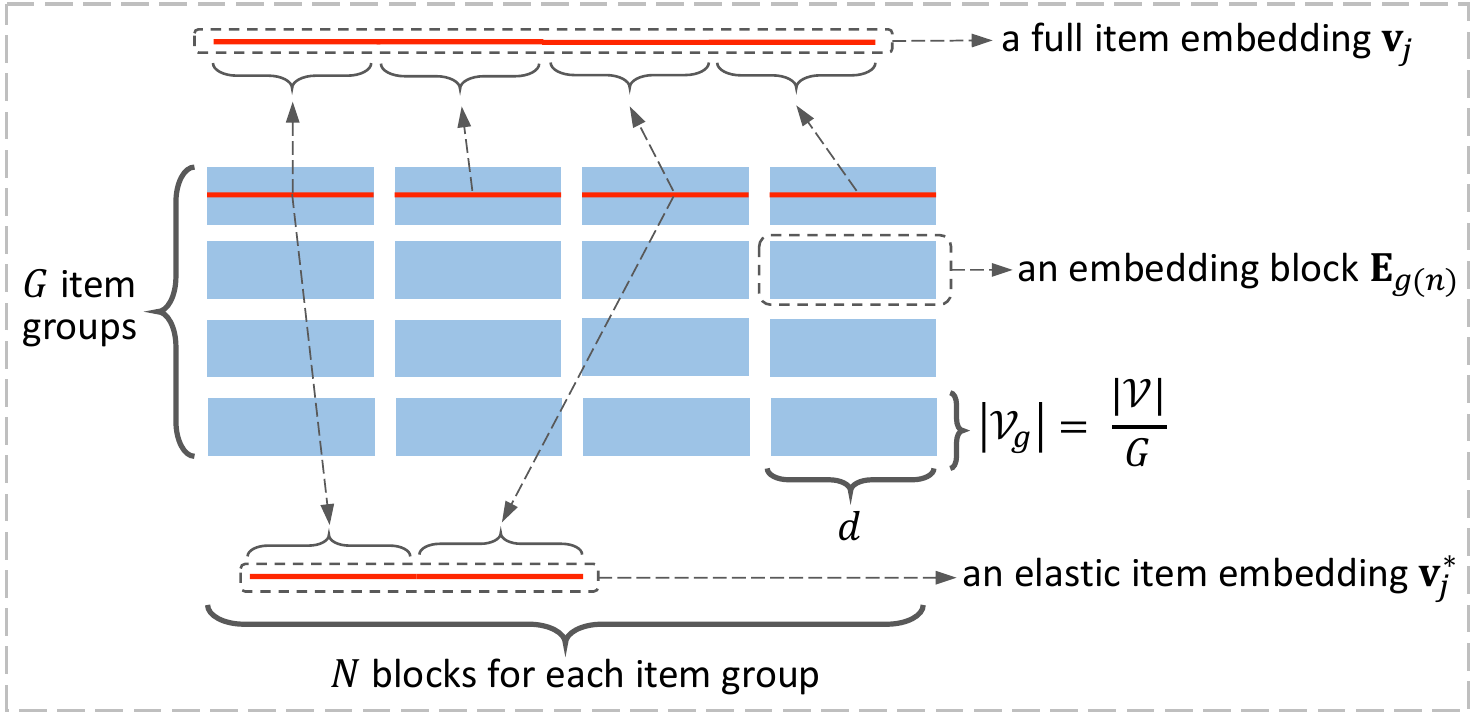}
\vspace{-0.4cm}
\caption{A graphic explanation on key notations used.}
\label{Figure:notations}
\end{figure}

Figure \ref{Figure:notations} provides a graphic view of the key notations w.r.t. the search space. The resulted search space has some desirable characteristics. Firstly, with $G \ll |\mathcal{V}|$, we can effectively reduce the number of embedding blocks to $G\times N$. Secondly, assuming items in each group share similar properties (e.g., items in the same category or cluster), then the selection of the best embedding blocks becomes a collective learning goal \cite{silva2020meteor}, where $\mathcal{B}_g$ is searched for optimizing the recommendation within $\mathcal{V}_g$ as a whole. As the strategies for segmenting items into groups will shape the search space of RULE, we will discuss and examine different strategies via experiments in Section \ref{sec:exp}. Recall that each device only needs to store one user embedding which has negligible size compared with all item embeddings, we formulate our problem below.

\noindent\textbf{Problem 1} (On-Device Recommendation with Customized Elastic Embeddings): For each on-device memory budget $M$ (measured by MB), RULE searches $\{\mathcal{B}_g\}_{g=1}^G$ for all item groups, providing accurate recommendations with $size(\{\textbf{v}^*_j\}_{j=1}^{|\mathcal{V}|} \cup \{\textbf{u}_i\}) \leq M$ where $\textbf{u}_i \in \mathbb{R}^D$ is an arbitrary user's full embedding.

\section{Learning The Full Embeddings}\label{sec:full_emb_learning} 

\vspace{-0.1cm}
\subsection{Base Recommender}
In RULE, before optimizing its memory footprint, we need to obtain expressive embeddings to guarantee high-quality recommendations. To achieve this, any latent factor model can be a good fit, ranging from matrix factorization \cite{koren2009matrix} to deep models \cite{he2017neuralcol}. In this paper, we utilize a graph neural network, namely the LightGCN \cite{he2020lightgcn} that recently advances the performance over various recommenders. Note that our main contribution lies in the elastic, memory-efficient recommendation framework rather than the base recommender, hence we describe it in a brief streak.

\textbf{Recommender Structure Summary.} Treating each user/item as a distinct node, the user-item interactions can be formulated as a bipartite graph, where we use $\mathcal{G}(u_i)$ and $\mathcal{G}(v_j)$ to respectively denote the one-hop neighbor sets of user $u_i$ and item $v_j$. Given an arbitrary user/item, its embedding is updated by propagating its neighbors' embeddings into $u_i$/$v_j$:
\vspace{-0.2cm}
\begin{equation}
	\textbf{v}^{(l)}_j =\!\!\!\! \sum_{i \in \mathcal{G}(v_j)}\eta_{ij}\textbf{u}^{(l-1)}_{i},\,\,\,\,\,\,
	\textbf{u}^{(l)}_i =\!\!\!\! \sum_{j \in \mathcal{G}(u_i)}\eta_{ij}\textbf{v}^{(l-1)}_{j},
\vspace{-0.1cm}
\end{equation}
where the user/item embeddings at the $l$-th layer ($l \geq 1$) is formed by aggregating all its neighbors' $D$-dimensional embeddings at the previous layer. $\eta_{ij} = (\sqrt{|\mathcal{G}(u_i)|\cdot|\mathcal{G}(v_j)|})^{-1}$ is a graph Laplacian normalization term that is proven effective in a variety of literatures~\cite{he2020lightgcn,kipf2017semi}. $\textbf{u}^{(0)}_i$ and $\textbf{v}^{(0)}_j$ are randomly initialized and will be trained via back-propagation. For notation simplicity, we denote the output embeddings from the \textbf{\textit{final layer}} $L$ as $\textbf{u}_i\!=\!\textbf{u}^{(L)}_i\!\in\! \mathbb{R}^{D}$ and $\textbf{v}_j\!=\![\textbf{e}_{j(1)}, \textbf{e}_{j(2)},...,\textbf{e}_{j(N)}]\!=\!\textbf{v}^{(L)}_j\!\in\! \mathbb{R}^{D}$, which are used in the subsequent recommendation and elastic embedding search phases. 

\vspace{-0.2cm}
\subsection{Handling Item Embeddings with Elastic Dimensions}
This is where our own design emerges, starting with a flexible pairwise scoring function that can adapt to item embeddings with elastic dimensions during inference. Generally, to rank all items in $\mathcal{V}$ for each $u_i$, a pairwise scoring function that estimates the similarity between a user-item embedding pair $(\textbf{u}_i, \textbf{v}_j)$ is indispensable. Specifically, distance metrics like dot product and cosine similarity are the most commonly used ones owing to their simplicity and efficacy. Those metrics work well when user and item vectors are of the same dimension, which is true during training (i.e., $\textbf{u}_i$, $\textbf{v}_j\in \mathbb{R}^{D}$). However, when elastic embedding $\textbf{v}_j^*\in \mathbb{R}^{|\mathcal{B}_g|d}$ ($|\mathcal{B}_g|d\leq D$) is in use for item $v_j$, methods like dot product become ill-posed once there is a mismatch between two embedding dimensions. Another common pairwise scoring scheme is to pass the concatenation of user and item embeddings into a multi-layer perceptron (MLP) with a linear regression layer~\cite{he2017neuralcol}. Though this makes two embeddings with unequal dimensions comparable, it is impractical by design as the MLP must be trained with a fixed input dimension (i.e., $2D$ in our case). So, for item $v_j \in \mathcal{V}_g$, we propose a dimension-independent scoring function below:
\vspace{-0.1cm}
\begin{equation}\label{eq:ranking_score}
	r_{ij} = \frac{\max(\{|\mathcal{B}_{g'}|\}_{g'=1}^{G})}{|\mathcal{B}_g|} \cdot \sum_{\forall n \in \mathcal{B}_g}\textbf{u}_i^{\top} \overline{\textbf{e}}_{j(n)},
\end{equation}
where $\overline{\textbf{e}}_{j(n)} = \concat_{\beta = 1,...,N}\textbf{e}_{j(n)} \in \mathbb{R}^D$ is constructed by repeating the same embedding block $\textbf{e}_{j(n)}$ for $N$ times. Noticeably, for all $\textbf{e}_{j(n)} \in \textbf{v}_j^*$ ($v_i \in \mathcal{V}_g$), we can result in $|\mathcal{B}_g|$ similarity scores. As different elastic item embeddings have varied numbers of embedding blocks, directly adding up all these scores might lead to significant variance in scale. Hence, we use the proportion between the maximum number of embedding blocks $\max(\{|\mathcal{B}_{g'}|\}_{g'=1}^{G})$ and the number of $\textbf{v}_j^*$'s embedding blocks $|\mathcal{B}_g|$ as a normalization term, making all user-item affinity scores $r_{ij}$ mutually comparable. Meanwhile, in the training process where the full item embedding $\textbf{v}_j$ is used, the normalization term is constantly $1$ and can be omitted. Also, this parameter-free design prevents the scoring function from stealing the memory budget from elastic embeddings.

\vspace{-0.2cm}
\subsection{Learning Diversified Embedding Blocks} 
To train RULE towards the recommendation task, we optimize all embeddings with Bayesian personalized ranking (BPR) \cite{rendle2009bpr} loss. At the same time, as each item's elastic embedding is composed by several different embedding blocks, the embedding blocks should be distinct from each other regarding the information they encode, thus maximizing the embeddings' expressiveness. Hence, on top of the recommendation loss, we further impose regularization effect on the diversity of all embedding blocks to be learned. The loss function of RULE is formulated as the following:
\vspace{-0.2cm}
\begin{equation}\label{eq:diversity_loss}
	L_{rec} = - \!\!\!\!\!\!\!\!\!\!\!\sum_{(u_i, v_{j^+}, v_{j^-})\in \mathcal{D}_{rec}}\!\!\!\!\!\!\varrho(r_{ij^+} - r_{ij^-}) - \lambda \sum_{n=1}^{N}\sum_{n'= n+1}^{N}\!\!\!\!||\textbf{E}_{n} - \textbf{E}_{n'}||_F^{2},
\end{equation}
where the first and second terms are respectively the BPR loss and the diversity regularizer. The rationale of the BPR loss is that, for user $u_i$, the ranking score for a visited item $v_{j^+}$ should always be higher than the ranking score for an unvisited one $ v_{j^-}$. As such, a training sample is defined as a triple $(u_i, v_{j^+}, v_{j^-}) \in \mathcal{D}_{rec}$, and $\mathcal{D}_{rec}$ denotes the set of all training samples. In the regularization term, $||\cdot||_F^{2}$ denotes the squared Frobenius norm, and $\textbf{E}_n = \concat_{g=1}^{G}\textbf{E}_{g(n)} \in \mathbb{R}^{|\mathcal{V}|\times d}$ is the collection of the $n$-th embedding block for all $G$ groups. It generally encourages two different embedding blocks to encode different information element-wise, thus posing a regularization effect on the diversity of learned embedding blocks.

\vspace{-0.2cm}
\section{Elastic Embedding Search}
Our RULE model can be viewed as an instantiated variant of automated neural architecture search (NAS). A commonly adopted workflow when deploying NAS for recommendation can be summarized as the iteration of two steps: (1) randomly generate candidate models (i.e.,  elastic embeddings in our case); and (2) train and evaluate the candidate models for the given task to find the currently best one. In RULE, training is exempted in step (2) as all elastic embeddings directly use the embedding blocks learned in Section \ref{sec:full_emb_learning}. But unlike most NAS-driven recommenders \cite{liu2020automated,zhao2020autoemb,song2020towards,wang2020autorec} whose objective only concerns the recommendation accuracy, in RULE, the searched elastic embeddings must also be memory-bounded so as to comply with the memory budget $M$. 

A straightforward solution is to formulate the problem as a reinforcement learning (RL) task \cite{joglekar2020neural}, where a trainable controller takes over step (1), and a performance reward computed in step (2) binds a memory cost term to demote the controller's selection of over-budget embedding block combinations.  
However, as the performance evaluation in step (2) is considerably time-consuming, it is prohibitively uneconomical to prepare a recommender for every possible on-device memory budget $M$. Also, most memory costs used in the reward are merely a ``soft'' limiter on the memory consumption \cite{joglekar2020neural}, which can be problematic for on-device applications if the size of elastic embeddings are not strictly bounded by $M$. Apart from RL, gradient-based NAS methods like DARTS~\cite{liu2019darts} also suffer from both defects in our recommendation paradigm.

In this section, we present our own solution to automated elastic embedding search. To bypass the underlying inflexibility while imposing strict memory constraints, we define a memory-bounded randomizer for generating candidate models, and perform evolutionary search with a pretrained performance estimator to construct the best-performing elastic embeddings. In what follows, we will expand the corresponding technical details. 

\begin{algorithm}[!t]
\begin{spacing}{0.9}
\small
\caption{Memory-Bounded Randomizer, i.e., $random(\cdot)$}\label{Algorithm:random}
\begin{algorithmic}[1]
\State \textbf{Input:} $N$, $G$, memory budget $M$ (converted to the total number of embedding blocks)
\State \textbf{Output:} Randomly selected embedding blocks $\{\mathcal{B}_g\}_{g=1}^{G}$
\State $model.EE\leftarrow \varnothing$, $\mathcal{S}\leftarrow \varnothing$, $\mu \leftarrow \frac{M}{G|\mathcal{V}_g|}$, $\sigma^2 \leftarrow 1$;
\For{$g\leftarrow 1,2,\cdots,G$} 
\State draw $s_g \!\sim\! \mathcal{N}(s|\mu, \sigma^2)$ where $s\!\in \!\{1,2,..., \textnormal{min}(N,M\!-G\!+\!1)\}$;
\State $\mathcal{S} \leftarrow \mathcal{S}\cup\{s_g\}$;
\EndFor
\State \textbf{end for}
\While{$\sum_{s \in \mathcal{S}}s >M$}
\State uniformly draw $g$ from $\{1,2,...,G\}$;
\If{$s_g>1$}
\State $s_g \leftarrow s_g -1$
\EndIf
\State \textbf{end if} 
\EndWhile
\State \textbf{end while}
\For{$g \leftarrow 1,2,\cdots,G$}
\State $\mathcal{B}_g \leftarrow$ draw $s_g$ unique embedding blocks from $\{\textbf{E}_{g(n)}\}_{n=1}^{N}$;
\State $model.EE \leftarrow model.EE \cup \{\mathcal{B}_g\}$;
\EndFor
\State \textbf{end for}
\State \Return{$model.EE$}
\end{algorithmic}
\end{spacing}
\end{algorithm}

\vspace{-0.3cm}
\subsection{Memory-Bounded Embedding Block Randomizer}
In RULE, we propose to enforce the memory constraint at an earlier stage, i.e., when randomly selecting embedding blocks for every $model.EE$. Here $model.EE$ denotes a candidate model's elastic embeddings. This is more computationally efficient as the selected embedding blocks fully comply with the memory budget, and the search for optimal elastic embeddings is now purely conditioned on the model performance. This is also achievable as the total number of embedding blocks that $model.EE$ will consume can be easily computed with any given $M$. For example, suppose $M=10$MB, $N=8$, $d=16$, $|\mathcal{V}|=100,000$, and $G=10$. Then, in a $32$-bit floating point system (i.e., $4\times 10^{-6}$MB per parameter), each embedding block occupies $0.64$MB memory from $\frac{d|\mathcal{V}|}{G}=1.6\times 10^5$ parameters. Considering the size of one user embedding $size(\textbf{u}_i) = 5.12\times 10^{-4}$MB, the maximum number of total embedding blocks that can fit in $M$ is $\lfloor \frac{M- size(\textbf{u}_i)}{0.64\textnormal{MB}} \rfloor=15$. 

Then, the problem comes down to deciding how many embedding blocks each item group will have, and which embedding blocks will be chosen. To further reduce the search complexity, we introduce a Gaussian prior, specifically standard normal distribution when determining the number of embedding blocks assigned to each group. The rationale is that, items' influence tend to follow normal distribution in recommendation tasks \cite{zhao2017exploiting,liu2013learning}, which can be also reflected via the distribution of their embedding dimensions. In other words, only the minority among items will be excessively impactful or irrelevant to the recommendation results and need substantially more/less embedding blocks, while the influences of most items are similar, so do their embedding dimensions.

We provide a mathematical view of the memory-bounded randomizer $random(\cdot)$ via Algorithm~\ref{Algorithm:random}, where we keep using $M$ to denote the maximum number of embedding blocks to be succinct. Specifically, we firstly assign $s_g\in\{1,2,...,\textnormal{min}(N,M\!-G\!+\!1)\}$ to each group $\mathcal{V}_g$ based on standard normal distribution $\mathcal{N}(\mu, \sigma^2)$, where the mean value $\mu$ is inferred from $M$ and variance $\sigma^2$ is set to $1$ (lines 4-7). Then, we iteratively calibrate the total number of embedding blocks to ensure it complies with the memory budget $M$ (lines 8-13). Lastly, once the numbers of embedding blocks is settled for all groups, we uniformly sample $s_g$ embedding blocks from $\{\textbf{E}_{g(n)}\}_{n=1}^{N}$ for the $g$-th item group (lines 15-17).

\vspace{-0.2cm}
\subsection{Performance Estimator} 
Evolutionary search \cite{real2019regularized} has been widely used for progressively searching for optimal neural architectures on a given task. However, in our case, evaluating every randomly sampled $model.EE$ on a validation dataset will impede its efficiency. In real-life scenarios, customized models need to be retrieved for arbitrary devices within a short time (e.g., a download request). Hence, we advocate training an accurate performance estimator to provide quick feedback on recommendation quality. Given $\{\mathcal{B}_g\}_{g=1}^{G}$, we first construct a vectorized input for the estimation function as follows:
\vspace{-0.1cm}
\begin{equation}
\label{eq:input}
	\textbf{x}_g = \textbf{W}\cdot \!\!\!\!\!\!\! \underbrace{[ \textnormal{one-hot}(g), \textnormal{multi-hot}(\mathcal{B}_g)]}_\textnormal{\footnotesize multi-hot encoding for the $g$-th item group}\!\!\!\!\!\!, 
\vspace{-0.2cm}
\end{equation}
where $\textnormal{one-hot}(g)$ is the $G$-dimensional one-hot encoding of the $g$-th item group, and $\textnormal{multi-hot}(\mathcal{B}_g)$ returns the $N$-dimensional multi-hot encoding representing the indexes of the selected embedding blocks. $\textbf{W}$ is a weight matrix projecting the multi-hot input for the $g$-th group into a $d_0$-dimensional vector. Then, to thoroughly capture the combinatorial effect among inter-group elastic embeddings, we build our performance estimator by modelling factorized pairwise interactions~\cite{rendle2010factorization,chen2020sequence} between item groups:
\vspace{-0.2cm}
\begin{equation}
\vspace{-0.2cm}
\label{eq:estimator}
\begin{split}
	\widehat{y} &= estimate(\{\textbf{x}_g\}_{g=1}^{G})\\
	&= \textbf{w}^{\top}\cdot FFN(\!\! \underbrace{\sum_{g=1}^{G}\sum_{g'=g+1}^{G}\textbf{x}_g \odot \textbf{x}_{g'}}_\textnormal{\footnotesize interactions between groups}\!\!) + b,
\end{split}
\vspace{-0.2cm}
\end{equation}
where $\odot$ denotes the element-wise multiplication, scalar $\widehat{y}$ is the predicted recommendation performance (e.g., recall value), $FFN(\cdot)$ is a single-layer feed-forward network that outputs a $d_0$-dimensional vector summarizing the pairwise interactions among all item groups' input features $\textbf{x}_g$. Then, the encoded combinatorial effect of different groups' elastic embedding compositions is mapped to $\widehat{y}$ with projection weight $\textbf{w}\in \mathbb{R}^{d_0}$ and scalar bias $b$. 

\textbf{Training The Performance Estimator.} We quantify the difference between the predicted performance and the actual one via:
\vspace{-0.2cm}
\begin{equation}\label{eq:est_loss}
L_{est} =\!\!\!\!\!\!\!\!\!\!\!\!\!\!\!\! \sum_{(model.EE, y_{model.EE}) \in \mathcal{D}_{est}} \!\!\!\!\!\!\!\!\!\!\!\!\!\!\!(y_{model.EE}- \widehat{y}_{model.EE})^2,
\vspace{-0.1cm}
\end{equation}
where $\mathcal{D}_{est}$ contains all $(model.EE, y_{model.EE})$ ground truth tuples for training. The ground truth is obtained by randomly acquiring a certain amount of different elastic embedding compositions $model.EE$ (with no memory restrictions) and evaluate them via any commonly used evaluation metrics like recall and precision (implementation details are provided in the Appendix). By controlling the size of $\mathcal{D}_{est}$, we can achieve a trade-off between the prediction confidence and training time of our performance estimator.

\begin{algorithm}[!t]
\begin{spacing}{0.9}
\small
\caption{Performance Estimator-based Evolutionary Search}\label{Algorithm:search}
\begin{algorithmic}[1]
\State \textbf{Input:} $M$, $N$, $G$, three search coefficients $P$, $S$, $C$ 
\State \textbf{Output:} RULE model with searched elastic item embeddings 
\State $seed \leftarrow \varnothing$, $cache \leftarrow \varnothing$;
\While{$|seed|<P$} 
\State $model.EE \leftarrow random(M, N, G)$;
\State $model.ACC \leftarrow estimate(model.EE)$;
\State $seed \leftarrow seed \cup \{child\}$;
\State $cache \leftarrow cache \cup \{child\}$;
\EndWhile
\State \textbf{end while}
\For{$c \leftarrow 1,2,\cdots,C$}
\State $sample \leftarrow \varnothing$;
\State $sample \leftarrow $ draw $S$ models with replacement from $seed$;
\State $parent \leftarrow model$ s.t. $model.ACC$ is the highest in $sample$; 
\State $child.EE \leftarrow mutate(parent.EE)$;
\State $child.ACC \leftarrow estimate(model.EE)$;
\State $seed \leftarrow seed \cup \{child\}$;
\State $cache \leftarrow cache \cup \{child\}$;
\State remove $model$ with the lowest $model.ACC$ from $seed$;
\EndFor
\State \textbf{end for}
\State \Return{$model$ with the highest $model.ACC$ in $cache$}
\end{algorithmic}
\end{spacing}
\end{algorithm}

\vspace{-0.3cm}
\subsection{Evolutionary Search with $estimate(\cdot)$}
To bypass the time-consuming model evaluation process, we perform evolutionary search with well-trained performance estimator. Algorithm \ref{Algorithm:search} depicts the search process, where $model.ACC$ denotes each candidate model's accuracy. Specifically, our performance estimator-based evolutionary search maintains a set of $P$ models (denoted as $seed$) with random elastic embedding combinations, which are initialized with our memory-bounded randomizer $random(\cdot)$ based on the given memory budget $M$. The evolution lasts for a total of $C$ rounds (lines 10-19). In every iteration, we draw $S$ models from $seed$ and select the one with the highest estimated performance as $parent$. Then, the $parent$ model is modified into a $child$ version with our $mutate(\cdot)$ function. We design two mutation operations on $parent$: (1) swapping the embedding blocks between two random item groups; (2) reselecting a same amount of embedding blocks for one item group at random. Only one mutation operation will be executed with uniform probability at each iteration, and the resulted model $child$ will be evaluated by our performance estimator. The rationale is that, if we keep revising the currently best model $parent$, it is likely to obtain more accurate results. After $C$ evolutionary rounds, the best-performing model is retrieved from $cache$ that stores all $seed$ and $child$ models. 

In this way, once the full embeddings and performance estimator are well trained in RULE, whenever it is required to perform recommendation on any device, RULE can quickly adapt to the device's memory constraint by selecting the most suitable embedding blocks so as to generate accurate recommendations.

\begin{table}[t!]
\vspace{0.1cm}
\caption{Statistics of experimental datasets.}
\vspace{-0.5cm}
\renewcommand{\arraystretch}{0.8}
\setlength\tabcolsep{3.5pt}
\center
  \begin{tabular}{c c c c c}
    \toprule
    Dataset & \#User & \#Item & \#Interaction & Sparsity\\
    \hline
    Amazon-Book & 52,643 & 91,599 & 2,984,108 & 99.94\% \\
	Yelp2020 & 138,322 & 105,843 & 3,865,586 & 99.97\% \\
    \bottomrule
\end{tabular}
\label{table:Dataset}
\end{table}

\begin{table*}[t!]
\vspace{-0.6cm}
\caption{Recommendation results. Numbers in bold face are the best results for corresponding metrics.}
\vspace{-0.4cm}
\centering
\renewcommand{\arraystretch}{0.9}
\setlength\tabcolsep{1.1pt}
  \begin{tabular}{c|c|cc|cc|cc|cc|cc|cc}
    \hline
    \multirow{3}{*}{Dataset} & \multirow{3}{*}{Method} & \multicolumn{4}{c|}{5MB} & \multicolumn{4}{c|}{10MB} & \multicolumn{4}{c}{25MB}\\
    \cline{3-14}
    & & \multicolumn{2}{c|}{Recall$@K$} & \multicolumn{2}{c|}{NDCG$@K$} & \multicolumn{2}{c|}{Recall$@K$} & \multicolumn{2}{c|}{NDCG$@K$} & \multicolumn{2}{c|}{Recall$@K$} & \multicolumn{2}{c}{NDCG$@K$}\\
    \cline{3-14}
    & & $K=50$ & $K=100$ & $K=50$& $K=100$ & $K=50$ & $K=100$ & $K=50$ & $K=100$ & $K=50$ & $K=100$ & $K=50$ & $K=100$ \\
    \hline
    \multirow{7}{*}{Amazon-Book} & PMF & 0.02457 & 0.04301 & 0.02552 & 0.03169 & 0.02717 & 0.04921 & 0.02873 & 0.03525 & 0.02878 & 0.05195 & 0.03050 & 0.03848 \\
    & LightGCN & 0.05070 & 0.08578 & 0.08864 & 0.10193 & 0.06460 & 0.09300 & 0.10738 & 0.13292 & 0.06517 & 0.08931 & 0.10937 & 0.14894 \\
    & ESAPN & 0.02148 & 0.04454 & 0.03943 & 0.06377 & 0.02336 & 0.04969 & 0.05269 & 0.08795 & 0.02561 & 0.06193 & 0.05611 & 0.08979\\
    & AutoEmb & 0.02024 & 0.04215 & 0.03890 & 0.05949 & 0.02327 & 0.05146 & 0.05033 & 0.08581 & 0.02496 & 0.06010 & 0.05412 & 0.08792\\
    & DLRM-CE & 0.02331 & 0.04722 & 0.04218 & 0.06948 & 0.02711 & 0.06247 & 0.05817 & 0.09726 & 0.03280 & 0.08363 & 0.06755 & 0.10010\\
    & LightRec & 0.04562 & 0.07040 & 0.07419 & 0.08982 & 0.05185 & 0.08289 & 0.08277 & 0.09825 & 0.05956 & 0.08093 & 0.09890 & 0.13765\\
    \cline{2-14}
    & \textbf{RULE}& \textbf{0.05335} & \textbf{0.08785} & \textbf{0.09079} & \textbf{0.12261} & \textbf{0.06537} & \textbf{0.09416} & \textbf{0.12787} & \textbf{0.15785} & \textbf{0.06622} & \textbf{0.10161} & \textbf{0.12983} & \textbf{0.18165}\\
    \hline
    \hline
    \multirow{7}{*}{Yelp2020} & PMF & 0.00798 & 0.01726 & 0.01048 & 0.01537 & 0.08160 & 0.01900 & 0.01273 & 0.01861 & 0.09126 & 0.02013 & 0.01542 & 0.02394 \\
    & LightGCN & 0.01327 & 0.02435 & 0.01971 & 0.02670 & 0.01784 & 0.02823 & 0.02772 & 0.04219 & 0.01992 & 0.03441 & 0.03188 & 0.05398\\
    & ESAPN & 0.00713 & 0.01959 & 0.01372 & 0.01980 & 0.00806 & 0.02186 & 0.01452 & 0.02119 & 0.00903 & 0.02325 & 0.01543 & 0.02179\\
    & AutoEmb & 0.00765 & 0.01912 & 0.01321 & 0.01920 & 0.07838 & 0.02087 & 0.01417 & 0.02236 & 0.00875 & 0.02214 & 0.01503 & 0.02093\\
    & DLRM-CE & 0.01214 & 0.02753 & 0.01239 & 0.01756 & 0.01381 & 0.02992 & 0.01596 & 0.02232 & 0.01484 & 0.03098 & 0.01945 & 0.02712\\
    & LightRec & 0.00805 & 0.02314 & 0.01536 & 0.02372 & 0.01259 & 0.02146 & 0.01487 & 0.02233 & 0.01334 & 0.02524 & 0.02098 & 0.03475\\
    \cline{2-14}
    & \textbf{RULE}& \textbf{0.02007} & \textbf{0.03602} & \textbf{0.02217} & \textbf{0.03282} & \textbf{0.02181} & \textbf{0.03875} & \textbf{0.03448} & \textbf{0.05447} & \textbf{0.02288}& \textbf{0.03947}& \textbf{0.03520}& \textbf{0.05529} \\
    \hline   
    \end{tabular}
\label{table:recommendation}
\vspace{-0.5cm}
\end{table*}

\section{Experiments}\label{sec:exp}
We evaluate the performance of RULE under different on-device memory constraints. In particular, we aim to answer the following research questions (RQs) via experiments:
\begin{itemize}
	\item[\textbf{RQ1:}] Under certain memory constraints, how is the recommendation accuracy of RULE compared with other baselines?
	\item[\textbf{RQ2:}] What is the contribution of each key component of RULE?
	\item[\textbf{RQ3:}] How is the on-device usability, specifically latency of RULE when being deployed on heterogeneous devices?
	\item[\textbf{RQ4:}] How the hyperparameters affect the performance of RULE?
\end{itemize}

\vspace{-0.3cm}
\subsection{Datasets and Baseline Methods}
We use two benchmark datasets for evaluation, namely \textbf{Amazon-Book} and \textbf{Yelp2020}. The Amazon review data is originally crawled and made public by \cite{he2016ups}. We select Amazon-Book for our experiments as it is the largest dataset in the collection. Yelp is an open review platform for various businesses (e.g., restaurants and hotels). We use its public dataset released in 2020\footnote{https://www.yelp.com/dataset}, where businesses are viewed as items. 
Both datasets are publicly available and consist of million-scale user-item interactions. Following \cite{rendle2009bpr}, we filter out inactive users with less than 10 interacted items and unpopular items visited by less than 10 users. Their primary statistics are shown in Table~\ref{table:Dataset}. We split both datasets with the ratio of $70\%$, $10\%$ and $20\%$ for training, validation and test, respectively. 

We compare with six baselines from three categories. The first ones are mainstream recommenders with fix-sized embeddings, namely probabilistic matrix factorization (\textbf{PMF}) \cite{mnih2007probabilistic} and \textbf{LightGCN}~\cite{he2020lightgcn}. The second type is AutoML-based lightweight recommenders using multidimensional embeddings, which are \textbf{ESAPN} \cite{liu2020automated} and \textbf{AutoEmb} \cite{zhao2020autoemb}. The key difference between them is that ESAPN is optimized via RL while AutoEmb has a differentiable search process. Lastly, we have two compositional embedding-based methods. One is \textbf{DLRM-CE} \cite{shi2020compositional} that generates compositional embeddings with a quotient-remainder trick. The other is \textbf{LightRec}~\cite{lian2020lightrec} which composes item embeddings with parallel codebooks. 

\vspace{-0.2cm}
\subsection{Evaluation Protocols}\label{sec:eva_protocol}
We test all methods with three on-device memory budgets, i.e., $5$MB, $10$MB, and $25$MB. To ensure all tested methods fit in the given memory budget, we adjust the corresponding embedding dimensions according to $M$ so that the total size of: (1) the embedding parameters needed for representing all items and one user; and (2)~all weights and biases used for ranking, is right below $M$. We leverage two ranking metrics, namely recall at rank $K$ (Recall$@K$) and normalized discounted cumulative gain at rank $K$ (NDCG$@K$) that are widely adopted in recommendation research \cite{chen2020try,yin2016spatio,chen2020social,sun2020go}. We adopt $K=50,100$ where all items that are unvisited by each user are taken as negative samples for evaluation. In short, Recall$@K$ measures the ratio of the ground truth items that are present on the top-$K$ list, and NDCG$@K$ evaluates whether the model can rank the ground truth items as highly as possible.

\vspace{-0.3cm}
\subsection{Recommendation Performance (RQ1)}\label{sec:recperf}
\textbf{Implementation Notes.} We report the hyperparameter settings of RULE in the Appendix for reproducibility, which are tuned via grid search. The Appendix also covers details on how the performance estimator's training set $\mathcal{D}_{est}$ is constructed and how it is trained. Regarding the elastic embeddings, the optimal hyperparameters are $\{d=8, N=16, D=Nd=128, G=20\}$, whose impact will be further discussed in subsequent sections. By default, RULE randomly segments all items into $G$ groups, and two other segmentation strategies will be investigated in Section \ref{sec:ablation}.

We list the overall recommendation performance of all tested methods in Table \ref{table:recommendation}. The first observation we can draw from the results is that, RULE consistently outperforms all baselines with a significant margin ($p < 0.05$ in paired $t$-test w.r.t. with the best baseline in each column). Specifically, compared with methods that use compositional embeddings (i.e., DLRM-CE and LightRec), RULE shows advantageous performance with all memory budgets, especially when the memory budgets are relatively small (i.e., 5MB and 10MB). This showcases the privilege of our proposed elastic embedding paradigm, which learns a set of unique embedding blocks for each item rather than sharing meta-embedding vectors among  items. It is also worth mentioning that, with the automated search function, RULE is efficiently customized for three different memory budgets after one training cycle, while all other baseline methods are individually trained towards each memory constraint.

At the same time, as fix-sized embedding methods, PMF and LightGCN still demonstrate adequate performance under the on-device memory constraints. It is worth noting that, LightGCN, on which our method is based, yields competitive results on Amazon-Book. On the sparser Yelp2020 dataset, when the memory size is too small to allow for sufficient embedding dimensions (e.g., 5MB), its performance deteriorates heavily. In contrast, by actively selecting the most useful embedding block for the current memory setting, RULE is able to maintain its performance at a higher level. Surprisingly, in spite of the sophisticated design of the NAS-based methods ESAPN and AutoEmb, they cannot achieve satisfying recommendation results when the search space for embedding dimensions is constrained by a tight memory budget. Another possible reason is that both methods are originally specialized for rating prediction, hence inevitably facing difficulties generalizing to the ranking task.

\vspace{-0.5cm}
\subsection{Effectiveness of Model Components (RQ2)}\label{sec:ablation}
\begin{figure}[t!]
\centering
\begin{tabular}{ccc}
	\vspace{-0.15cm}\includegraphics[width=1.15in]{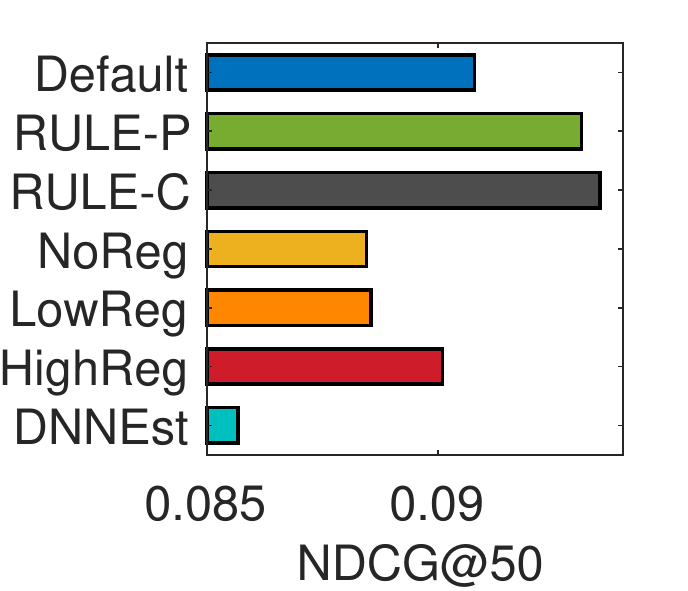}
	&\hspace{-0.4cm}\includegraphics[width=1.15in]{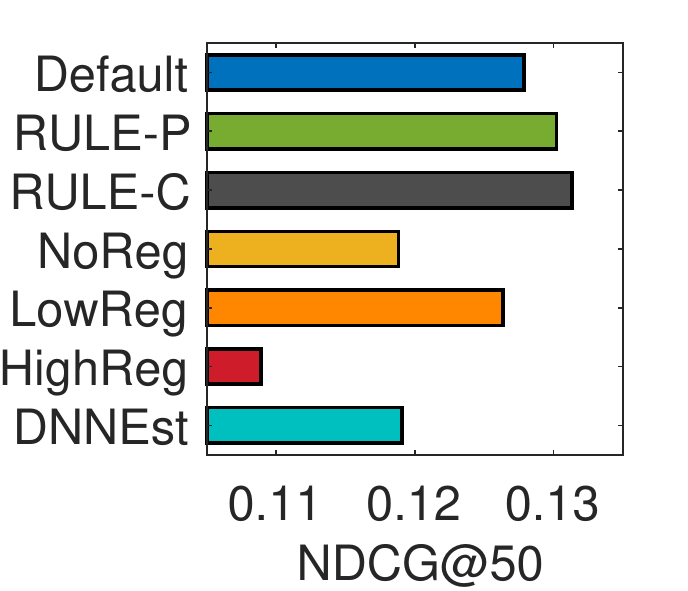}&\hspace{-0.5cm}\includegraphics[width=1.15in]{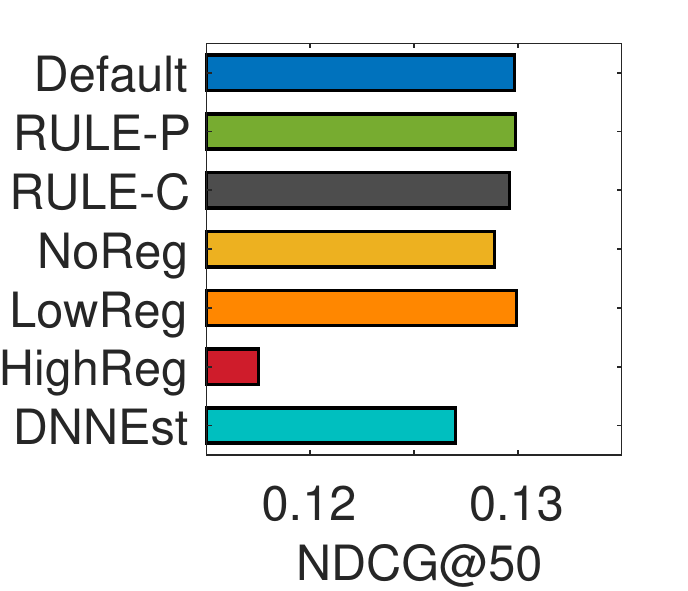}\\
	\multicolumn{3}{c}{\vspace{-0.05cm}\footnotesize{(a) Results on Amazon-Book within 5MB (left), 10MB (middle), and 25MB (right).}}\\
		\vspace{-0.15cm}\includegraphics[width=1.15in]{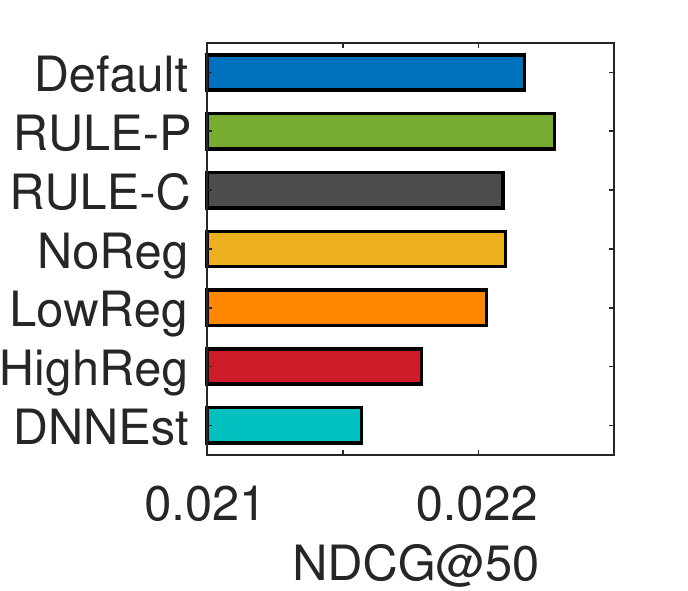}
	&\hspace{-0.4cm}\includegraphics[width=1.15in]{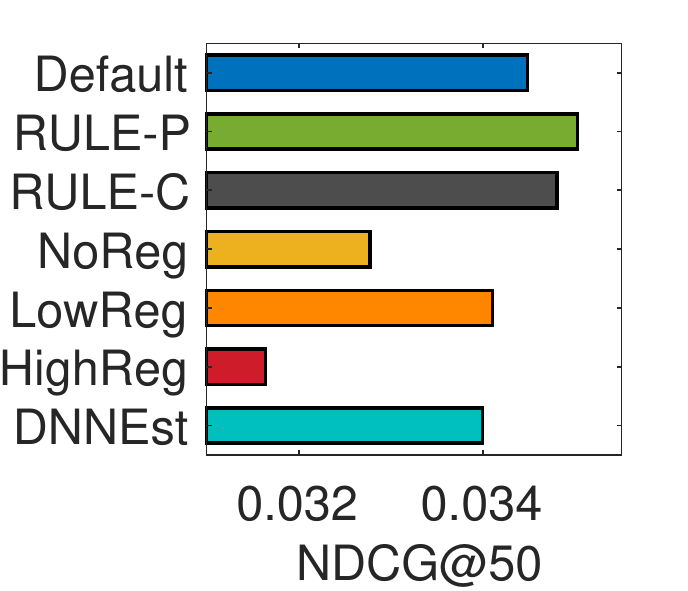}&\hspace{-0.5cm}\includegraphics[width=1.15in]{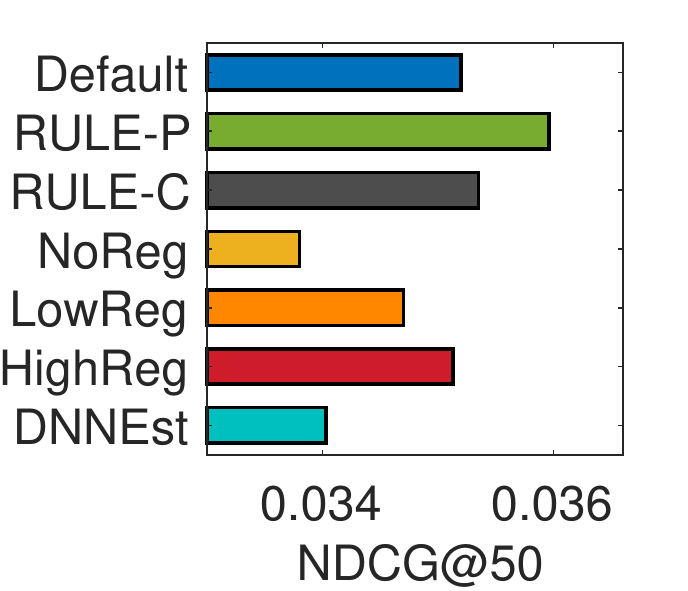}\\
	\multicolumn{3}{c}{\footnotesize{(b) Results on Yelp2020 within 5MB (left), 10MB (middle), and 25MB (right).}}\\
\end{tabular}
\vspace{-0.5cm}
\caption{Performance of different variants of RULE.}
\label{Figure:Ablation}
\end{figure}

To better understand the performance gain from different key components, we implement several variants of RULE by making modifications to one component each time, and record the new performance achieved. The results with all three memory budgets are reported via Figure \ref{Figure:Ablation}. We choose NDCG$@50$ as a performance indicator in this study. In what follows, we introduce all variants and analyze the effect of corresponding model components.

\begin{figure*}[t!]
\centering
\begin{minipage}[b]{.7\textwidth}
\hspace{-0.2cm}
\begin{tabular}{cc}
	\includegraphics[height=1in]{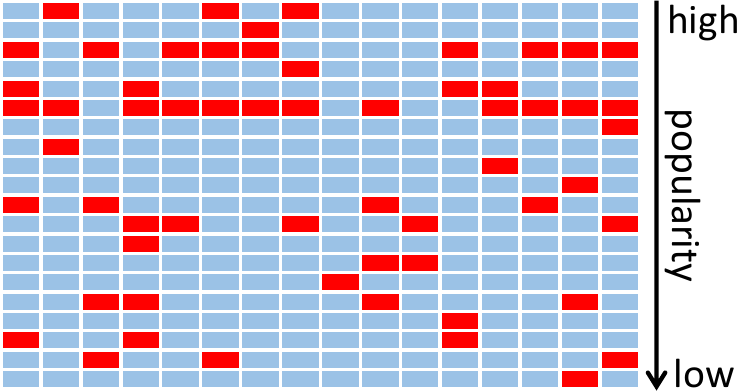}
	&\hspace{-0.2cm}\includegraphics[height=1in]{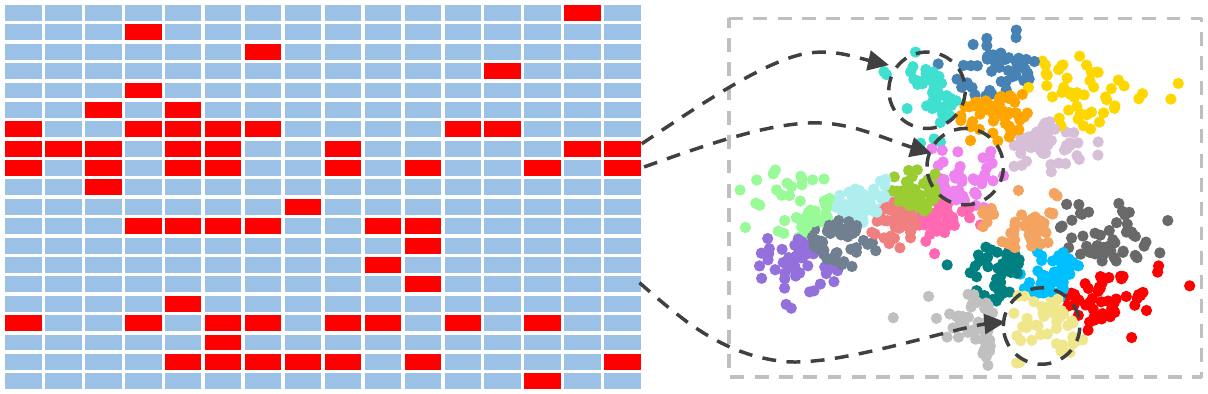}\\
	\footnotesize{(a) Popularity-based Item Groups} & \hspace{-0.5cm}\footnotesize{(b) Clustering-based Item Groups}\\
\end{tabular}
\vspace{-0.45cm}
\caption{Visualization of searched elastic embeddings with both popularity-based item groups and clustering-based item groups. The visualization corresponds to the Yelp2020 dataset under 10MB budget (59 embedding blocks searched). For better clarity, each cluster is downsampled to 50 items in (b).}\label{Fig:item_group}
\end{minipage}\qquad
\begin{minipage}[b]{.23\textwidth}
\hspace{3.5cm}\includegraphics[height=1in]{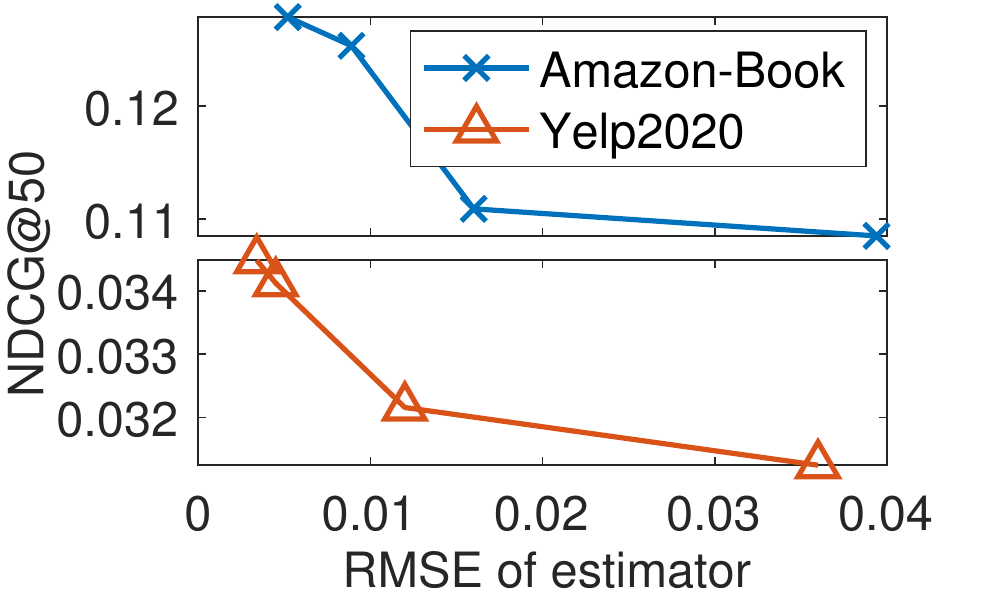}
\vspace{-0.75cm}
\caption{How the performance estimator affects the final recommendation accuracy on Yelp2020. 10MB budget is used for illustration.}\label{Fig:estimator}
\end{minipage}
\end{figure*}

\textbf{Different Item Segmentation Strategies.} We further testify two empirical strategies for defining item groups, namely popularity-based (RULE-P) and clustering-based item groups (RULE-C) with $G=20$ for both. More implementation details are provided in the Appendix. As the results suggest, by grouping items based on their common properties, the recommendation accuracy can be slightly improved in some cases. To further interpret the rationale of these two item segmentation strategies, we visualize the searched elastic embeddings in RULE-P and RULE-C in Figure \ref{Fig:item_group}. As can be told from Figure \ref{Fig:item_group}(a), popular item groups are more likely to receive more embedding blocks (i.e., larger embedding dimensions), thus fully encoding the rich information from interactions with users. In Figure \ref{Fig:item_group}(b), with the item clusters visualized, we empirically find that similar elastic embedding compositions can be found from two item groups if their clusters are close to each other, and vice versa. Hence, it might be of interest to practitioners to pursue better recommendation results with finer item segmentation strategies.

\textbf{Diversity-driven Regularizer.} To verify the contribution of the diversity-driven regularizer in Eq.(\ref{eq:diversity_loss}), we derive three variants of RULE by setting $\lambda$ to $0$, $1\times10^{-6}$ and $1\times10^{-2}$ (denoted by NoReg, LowReg, and HighReg in Figure \ref{Figure:Ablation}, respectively, and $\lambda = 1\times10^{-4}$ by default). Compared with LowReg, a slightly higher regularization weight as used in the default setting is more beneficial. When the regularizer is removed, the embedding blocks become less diversified, hence the composed elastic embeddings are less informative. Meanwhile, an overwhelmingly strong regularization will let the optimization process overlook the accuracy objective, impeding the final performance.

\textbf{FM-based Performance Estimator.} By replacing our factorization machine-based (FM-based) performance estimator with a single-layer deep neural network (denoted by DNNEst in Figure \ref{Figure:Ablation}) fed with the concatenation of all item groups' input vectors from Eq.(\ref{eq:input}), a performance drop is observed. This verifies that modelling the combinatorial effect between item groups leads to more accurate results on both estimated and actual recommendation performance. Furthermore, we quantitatively evaluate the association between the FM-based estimator and the final recommendation performance. For the FM-based performance estimator, we plot its training root mean square errors (RMSEs) at four different training stages and the corresponding recommendation results achieved on both datasets in Figure \ref{Fig:estimator}. Apparently, the actual recommendation performance of RULE is positively associated with the precision of function $estimate(\cdot)$ that directs the search process.

\begin{figure}[t!]
\vspace{-0.6cm}
\centering
\begin{tabular}{cc}
	\vspace{-0.1cm} \includegraphics[width=1.7in]{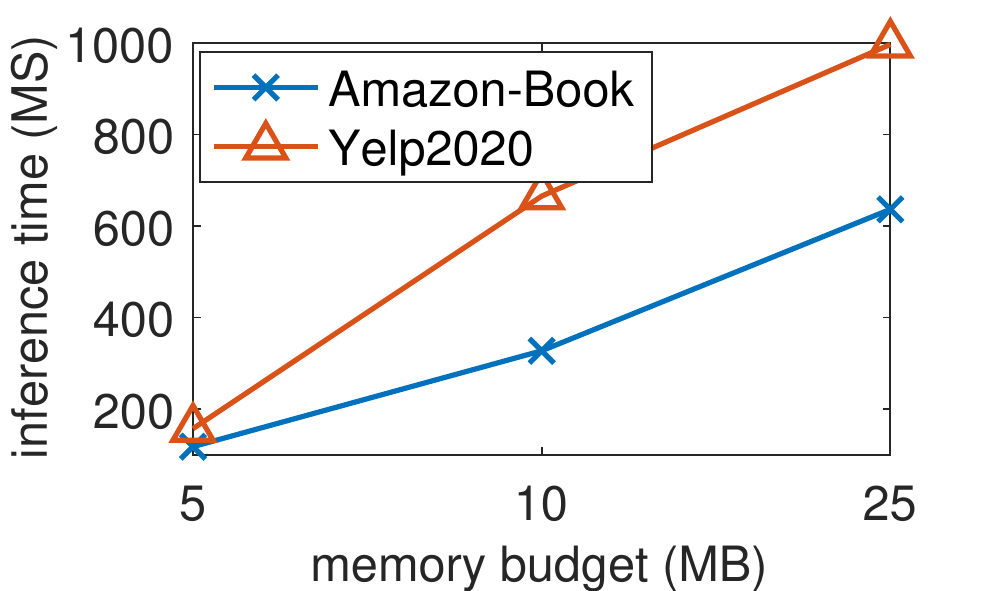}
	&\hspace{-0.5cm}\includegraphics[width=1.7in]{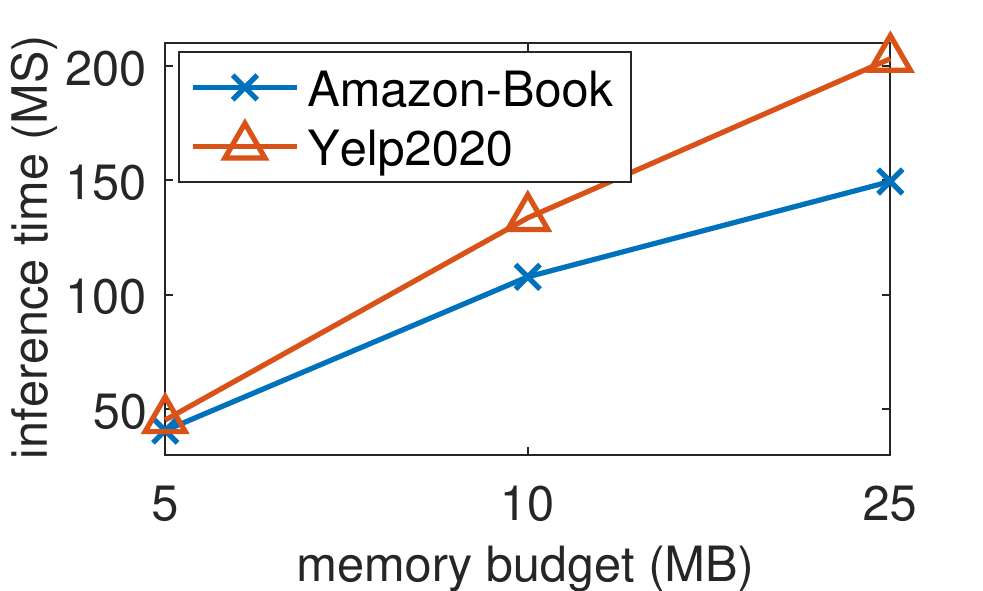}\\
	\footnotesize{(a) IoT (single-core CPU)} & \hspace{-0.5cm}\footnotesize{(b) Cellphone (four-core CPU)}\\
\end{tabular}
\vspace{-0.5cm}
\caption{Average inference time of RULE for each user, tested with two on-device environments.}\label{Fig:time}
\end{figure}

\vspace{-0.4cm}
\subsection{Latency on Heterogeneous Devices (RQ3)}
Since the main use case of our proposed recommendation paradigm is on-device services, we test the real-life practicality, specifically the  latency of RULE in two hardware configurations. With the help of a Linux virtual machine, we deploy RULE in simulated IoT and cellphone environments, and report the average inference time for generating a full ranked item list for each user. We describe the on-device settings and the latency of RULE below.

\textbf{IoT: $1\times$ vCPU (Intel i7-7900K), 64MB RAM.} Note that for both IoT and cellphone configurations, we choose a more realistic number for the RAM size rather than simply our largest memory budget (i.e., 25MB) due to real-world multitasking demands. As we can see from Figure \ref{Fig:time}(a), with a compact embedding size under 5MB, it takes less than 200MS for the item ranking list to be prepared. Even when $M$ stretches to 25MB on Yelp2020, RULE can still keep the inference time under the 1000MS mark.

\textbf{Cellphone: $4\times$ vCPU (Intel i7-7900K), 512MB RAM.} With three more computing cores, RULE has demonstrated minor latency when performing recommendation in a cellphone environment. When the memory budget increases from 5MB to 25MB, the execution time grows from under 50MS on both datasets to 148.5MS and 203.19MS on Amazon-Book and Yelp2020, respectively. To summarize, RULE is capable of learning lightweight elastic embeddings for a wide range of on-device recommendation tasks. 

\vspace{-0.4cm}
\subsection{Hyperparameter Analysis (RQ4)}
\begin{figure}[t!]
\vspace{-0.5cm}
\centering
\begin{tabular}{cc}
	\vspace{-0.1cm}\includegraphics[width=1.6in]{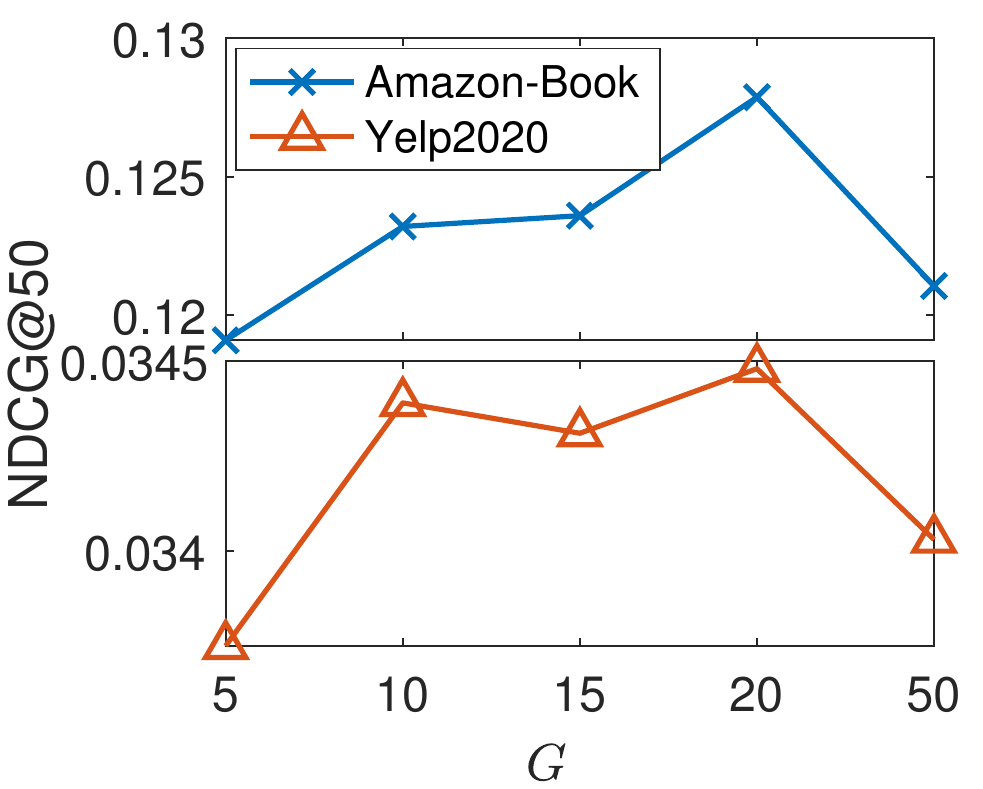}
	&\hspace{-0.2cm}\includegraphics[width=1.6in]{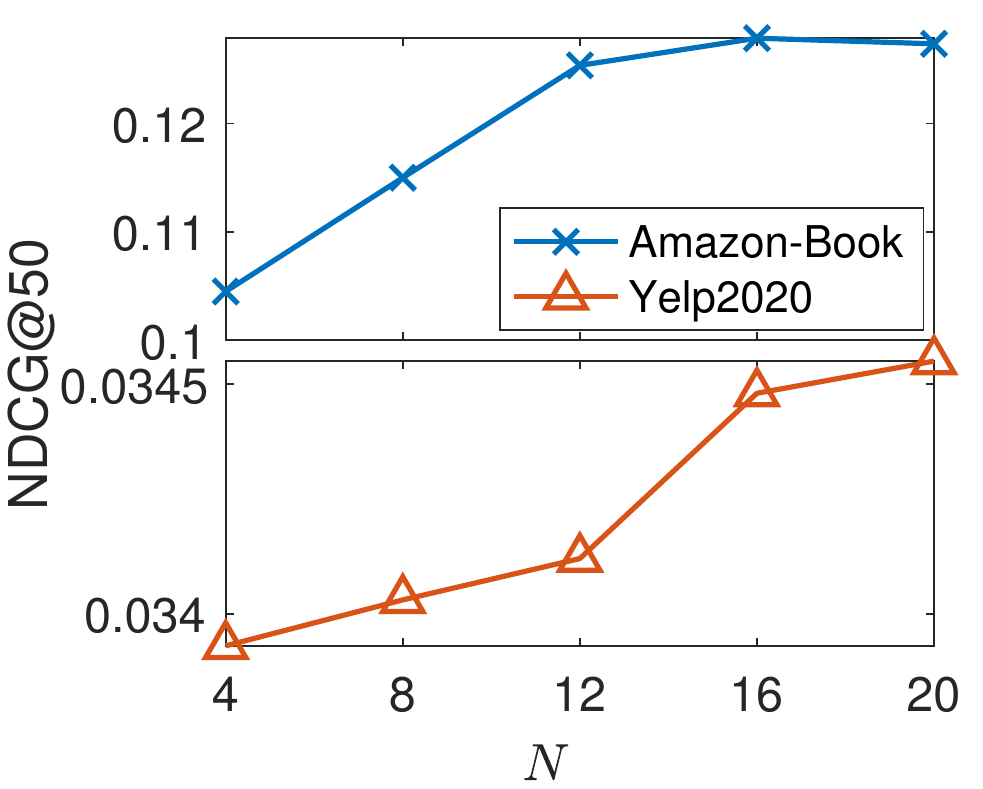}\\
	\footnotesize{(a) Performance of RULE w.r.t. $G$.} & \hspace{-0.3cm}\footnotesize{(b) Performance of RULE w.r.t. $N$.}\\
\end{tabular}
\vspace{-0.5cm}
\caption{Impact of $G$ and $N$ to the recommendation performance. Note that we set $M=$10MB for demonstration while similar trends are observed with other memory budgets.}
\label{Fig:hyperparam}
\end{figure}

We further study the impact of two key hyperparameters, namely the number of item groups $G$ and the number of embedding blocks $N$, to the performance of RULE. Specifically, we vary the value of $G$ or $N$ at a time while keeping other hyperparameters unchanged, and report the recommendation results achieved via Figure \ref{Fig:hyperparam}. Similar to Section \ref{sec:ablation}, we set $M=10$MB and use NDCG$@50$ for demonstration, and similar performance fluctuations are observed with other memory budgets.
 
\textbf{Impact of $G$.} We study the impact of group number with $G\in\{5,10,15,20,50\}$. The size of $G$ affects the granularity of the segmented item groups. As can be seen from Figure \ref{Fig:hyperparam}(a), better results are obtained when G ranges from $10$ to $20$. When the item groups are too coarse (i.e., $G=5$), it limits the potential of identifying the most suitable embedding blocks for all items. Meanwhile, with $G=50$, it creates a large search space, making it challenging for RULE to identify the optimal elastic embedding compositions.

\textbf{Impact of $N$.} We vary $N$ in $\{4,8,12,16,20\}$. Note that $d=8$ remains fixed, resulting in a full embedding dimension $D=Nd\in\{32, 64, 96, 128, 160\}$. As Figure \ref{Fig:hyperparam}(b) suggests, more embedding block candidates generally contribute to higher recommendation accuracy. However, as $N$ reaches $16$ and $20$, the performance increase becomes marginal because the number of embedding blocks that can be chosen is strictly controlled by the memory budget.

\vspace{-0.4cm}
\section{Related Work}\label{sec:related}
Recommendation models that are memory-efficient has gained immense attention over the years, owing to the increasing need for decentralization \cite{muhammad2020fedfast} and IoT-compatibility \cite{wang2020next}. 
Different from methods designed for tasks like neural translation and image processing \cite{kitaev2020reformer,cai2019once} where the excessive model parameters (i.e., weights and bias) are the main source of memory consumption, most recommender systems' parameterization concentrates on the embedding layers for discrete features, especially user/item IDs. In this regard, a straightforward solution is to convert all real-valued embedding vectors into fixed-length binary codes (e.g., discrete collaborative filtering \cite{lian2017discrete,zhou2012learning,zhang2018discrete,zhang2020deep}) so as to shrink the space needed for storing them. However, it is well-known that the binarization of real-valued parameters will lead to significant performance drop due to the quantization loss \cite{liu2018discrete,zhang2016discrete}, making discrete methods less ideal for recommendation. Though deep neural networks are later applied as an enhancement upon the binary user/item codes \cite{zhang2018discrete}, discrete methods inherently falls short in generating semantically rich user/item representations, and hence face an inevitable compromise in recommendation accuracy. 

Recently, another variant of lightweight recommendation models are devised based on compositional embeddings \cite{lian2020lightrec,shi2020compositional,chen2020differentiable}, some of which are inspired by word embedding compression in language modelling \cite{suzuki2016learning,shu2018compressing,panahi2019word2ket}. The key idea of compositional embeddings is to quantize a full user/item embedding matrix into a smaller one with substantially less embedding vectors (a.k.a. meta-embeddings). In such embedding systems, each user/item is represented as a learned composition of several meta-embeddings from the small embedding matrix, such that the embeddings used for recommendation are still highly distinctive and significant parameter reduction can be achieved. In~\cite{wang2020next}, tensor-train decomposition is adopted to compress embeddings, which can also be interpreted as a specific quantization operation applied to each element of the full embedding table \cite{shi2020compositional}. However, as pointed out in Section \ref{sec:intro}, those methods are all designed towards a predefined memory constraint. When being applied to heterogeneous devices, a model have to be rebuilt and retrained in order to comply with a different memory budget. In summary, the pursuit of lightweight recommender systems that are adaptive to heterogeneous devices is still underway.

As recommender systems start to benefit from memory efficiency in contemporary real-life applications, there have also been emerging research efforts \cite{joglekar2020neural,liu2020automated,zhao2020autoemb} on developing lightweight recommendation models with neural architecture search (NAS). The core idea is to reduce the memory footprint of embeddings by searching for optimal embedding dimensions of each object (e.g., item, user, feature, etc.) instead of using fixed embedding dimensions for all. In \cite{liu2020automated,zhao2020autoemb,zhao2020memory}, the search space is defined by assigning each object a set of embeddings with different dimensions, e.g., learning 3 embeddings for each item with dimensionality of 32, 64, and 128. To avoid over-parameterization of the original embedding space, an alternative definition of the search space is utilized in~\cite{joglekar2020neural} where a full embedding matrix can be sliced into multiple embedding blocks. Then, automated optimization strategies based on either reinforcement learning or DARTS are respectively adopted by \cite{liu2020automated,joglekar2020neural} and \cite{zhao2020memory,zhao2020autoemb} in order to determine the most suitable embedding size for each object. Unfortunately, after a long training and search process, existing NAS-based recommenders can only produce one final model, making them unsustainable for real-time deployment with heterogeneous memory budgets on different devices during. In contrast, our proposed RULE enables an ``once-for-all'' recommendation paradigm that can efficiently scale up to heterogeneous on-device memory constraints.

\vspace{-0.2cm}
\section{Conclusion}\label{sec:conclusion}
In this paper, we propose a new paradigm for on-device recommendation, which automatically customizes elastic item embeddings for different memory constraints without retraining. With our novel solution RULE, we firstly learn diversified yet informative embedding blocks, then design a performance estimator-based evolutionary search method to efficiently identify the optimal elastic embedding composition for each item group. The resulted elastic item embeddings fully comply with the strict on-device memory budget, and experiments verify that RULE offers strong guarantees on the recommendation performance and efficiency.

\section*{Acknowledgement}
This work is supported by the Australian Research Council under the streams of Discovery Project (No. DP190101985), Training Centre (No. IC200100022), and Centre of Excellence (No. CE200100025).


\clearpage
\newpage
\section*{Appendix on Reproducibility}
\appendix
\section{Hyperparameters of RULE}
As discussed in Section \ref{sec:recperf}, the hyperparameters in RULE are optimized via grid search. Table \ref{table:hyperparams} provides details on the optimal parameters used in RULE, as well as their corresponding search intervals. Unless specified, we use the following hyperparameters as the default settings of RULE in our experiments.
\begin{table}[h!]
\caption{Hyperparameter settings.}
\vspace{-0.5cm}
\renewcommand{\arraystretch}{1.0}
\setlength\tabcolsep{2pt}
\center
  \begin{tabular}{c c c c}
    \toprule
    \multirow{2}{*}{Dataset} & Hyper- & \multirow{2}{*}{Value} & \multirow{2}{*}{Search Interval}\\
    & parameter & &\\
    \hline
    \multirow{8}{*}{Amazon-Book} & $L$ & 3 & \{1,2,3,4,5\} \\
	 & $(N,d)^*$ & (16,8) & \{(4,8), (8,8), (12,8), (16,8), (20,8)\} \\
	 & $G$ & 20 & \{5,10,15,20,50\} \\
	 & $\lambda$ & 1$\times$10$^{-4}$ & \{0, 1$\times$10$^{-2}$, 1$\times$10$^{-4}$, 1$\times$10$^{-6}$\} \\
	 & $d_0$ & $64$ & \{8,16,32,64,128\} \\
	 & $P$ & $20$ & \{10,20,30,40,50\} \\
	 & $C$ & $50$ & \{10,20,30,40,50\} \\
	 & $S$ & $5$ & \{2,5,10,15,20\} \\
    \hline
    \multirow{8}{*}{Yelp2020} & $L$ & 2 & \{1,2,3,4,5\} \\
	 & $(N,d)^*$ & (16,8) & \{(4,8), (8,8), (12,8), (16,8), (20,8)\} \\
	 & $G$ & 20 & \{5,10,15,20,50\} \\
	 & $\lambda$ & 1$\times$10$^{-4}$ & \{0, 1$\times$10$^{-2}$, 1$\times$10$^{-4}$, 1$\times$10$^{-6}$\} \\
	 & $d_0$ & $128$ & \{8,16,32,64,128\} \\
	 & $P$ & $50$ & \{10,20,30,40,50\} \\
	 & $C$ & $50$ & \{10,20,30,40,50\} \\
	 & $S$ & $5$ & \{2,5,10,15,20\} \\
	 \bottomrule
	 \multicolumn{4}{l}{$^*D$ is determined via $D=Nd$}
\end{tabular}
\label{table:hyperparams}
\end{table}

\section{Training The Performance Estimator}
An accurate performance estimator is crucial to the quality of searched elastic embeddings. After finish learning the full user and item embeddings as described in Section \ref{sec:full_emb_learning}, we build the training dataset $\mathcal{D}_{est}$ for $estimate(\cdot)$. To obtain $(model.EE, y_{model.EE})$ pairs, we randomly sample $\beta =15,000$ elastic embedding compositions, and evaluate their performance on the validation set. To let RULE easily generalize to different memory budgets without retraining, the $\beta$ different elastic embedding compositions are sampled without any predefined memory budget. Then, each sampled $model.EE$ will be evaluated on the same set of $\mu=1,000$ users from the validation set to obtain $y_{model.EE}$. 
To measure the performance of each $model.EE$, the performance metric we choose is Recall$@100$, however this is fully customizable for other purposes (e.g., using AUC for click-through rate prediction). Note that we evaluate sampled elastic embeddings on $\mu$ users instead of all users because this will substantially reduce the time consumption, while still ensuring each resulted $y_{model.EE}$ is a high-quality indicator on the recommendation performance. With the randomizer (i.e., $random(\cdot)$) defined in Algorithm \ref{Algorithm:random}, the construction process of $\mathcal{D}_{est}$ is described in Algorithm \ref{Algorithm:D}.

With the training set $\mathcal{D}_{est}$, we train our performance estimator with the loss defined in Eq.(\ref{eq:est_loss}). At the initial training stage, we perform two-fold cross-validation on $\mathcal{D}_{est}$ when determining its hyperparameters. After the hyperparameters are set, we retrain the estimator with the full set $\mathcal{D}_{est}$ until the loss converges.

\begin{algorithm}[!h]
\begin{spacing}{0.9}
\small
\caption{Construction Process of $\mathcal{D}_{est}$}\label{Algorithm:D}
\begin{algorithmic}[1]
\State \textbf{Input:}  $N$, $G$, $\mu$, $\beta$
\State \textbf{Output:} $\mathcal{D}_{est}$
\State $\mathcal{D}_{est}\leftarrow \varnothing$;
\For{$b\leftarrow 1,2,\cdots,\beta$} 
\State draw $\widetilde{M}$ uniformly at random from $\{G, G\!+\!1, ...,NG\}$;
\State $model.EE \leftarrow random(\widetilde{M}, N, G)$;
\State $y_{model.EE} \leftarrow$ evaluate model.EE with Recall$@100$ on the same $\mu$ 
\Statex \hspace{2.1cm} validation users;
\State $\mathcal{D}_{est}\leftarrow(model.EE, y_{model.EE})$;
\EndFor
\State \textbf{end for}
\State \Return{$\mathcal{D}_{est}$}
\end{algorithmic}
\end{spacing}
\end{algorithm}

\section{Item Segmentation Strategies}
By default, RULE segments all items randomly into $G$ groups. In Section \ref{sec:ablation}, we further testify two more item segmentation strategies, i.e., popularity-based and clustering-based segmentation. We present more implementation details as follows.

\textbf{Popularity-based Segmentation.} It is widely acknowledged that item popularity plays an important role in recommendation~\cite{liu2020automated}, which motivates our practice of discriminatively treating items with varied popularity. Intuitively, the embedding size determines the capacity to encode information. As popular items tend to have abundant interaction records for training, a larger embedding dimension (i.e., more embedding blocks) is helpful for comprehensively capturing contexts associated with them. On the contrary, less embedding blocks are needed when representing long-tail items that offer limited predictive signals. So, we sort all items $v_j\in \mathcal{V}$ in descending order according to their popularity (i.e., times interacted by different users), such that $v_1$ and $v_{|\mathcal{V}|}$ are respectively the most and least popular items. Then, $G$ item groups can be formed by evenly segmenting the sorted item set. 

\textbf{Clustering-based Segmentation.} As the search space only needs to be constructed after all items' full embeddings $\textbf{v}_j = [\textbf{e}_{j(1)}, \textbf{e}_{j(2)},$ $...,\textbf{e}_{j(N)}]$ are learned, another natural strategy for segmenting item groups is to cluster items based on their finely trained full embeddings. Specifically, we leverage $k$-Means that is a scalable and well-established algorithm to map all items into $G$ clusters. With principal component analysis (PCA), we first convert all items' full embeddings into 2-dimensional vectors. Then, by setting $k=G=20$ and applying $k$-Means on all 2-dimensional item embeddings, we can obtain $20$ clusters, which are treated as item groups in RULE. Note that we apply PCA prior to $k$-Means to facilitate our 2-dimensional visualization in Figure \ref{Fig:item_group}, and this step is optional for normal use. Essentially, items falling in the same cluster are more likely to exhibit close properties, making it ideal for them to share the same elastic embedding compositions. 

\end{document}